\documentclass[pdflatex,iicol,sn-mathphys-num]{sn-jnl}

\usepackage{graphicx}%
\usepackage{multirow}%
\usepackage{amsmath,amssymb,amsfonts}%
\usepackage{amsthm}%
\usepackage{mathrsfs}%
\usepackage[title]{appendix}%
\usepackage{xcolor}%
\usepackage{textcomp}%
\usepackage{manyfoot}%
\usepackage{booktabs}%
\usepackage{algorithm}%
\usepackage{algorithmicx}%
\usepackage{algpseudocode}%
\usepackage{listings}%
\usepackage{lmodern}

\theoremstyle{thmstyleone}%
\theoremstyle{thmstyletwo}%

\theoremstyle{thmstylethree}%

\begin{document}
\title[Article Title]{DipMe: Haptic Recognition of Granular Media for Tangible Interactive Applications}

\author[1]{\fnm{Xinkai} \sur{Wang}}\email{xkwang@seu.edu.com}

\author[2]{\fnm{Shuo} \sur{Zhang}}

\author[1]{\fnm{Ziyi} \sur{Zhao}}
\author[1]{\fnm{Lifeng} \sur{Zhu}}\email{lfzhulf@outlook.com}

\author[1]{\fnm{Aiguo} \sur{Song}}

\affil[1]{\orgdiv{School of Instrument Science and Engineering}, \orgname{Southeast University}, \orgaddress{\street{Sipailou No.2}, \city{Nanjing}, \postcode{210096}, \state{Jiangsu}, \country{China}}}

\affil[2]{\orgdiv{School of Artificial Intelligence}, \orgname{Jilin University}, \orgaddress{\street{Xinmin Road}, \city{Changchun}, \postcode{130021}, \state{Jilin}, \country{China}}}



\abstract{
While tangible user interface has shown its power in naturally interacting with rigid or soft objects, users cannot conveniently use different types of granular materials as the interaction media. 
We introduce \emph{DipMe} as a smart device to recognize the types of granular media in real time, which can be used to connect the granular materials in the physical world with various virtual content.
Other than vision-based solutions, we propose a dip operation of our device and exploit the haptic signals to recognize different types of granular materials. With modern machine learning tools, we find the haptic signals from different granular media are distinguishable by \emph{DipMe}. 
With the online granular object recognition, we build several tangible interactive applications, demonstrating the effects of \emph{DipMe} in perceiving granular materials and its potential in developing a tangible user interface with granular objects as the new media. }




\maketitle

\section{Introduction}\label{sec1}

Granular materials are commonly seen in our daily lives. As a continuous deformable media, granular materials such as sand or beads has been introduced to tangible user interface (TUI) to preview the landscape or adjust the stiffness of input devices \cite{ishii2004bringing,sandcanvas,jamming,sandscape}. 
While different rigid objects have been used in TUI to give flexible control of virtual content \cite{boem2019non,hashimoto2005grasping,peschke2012depthtouch,sato2009photoelastictouch,troiano2015deformable}, to the most of our knowledge, the potential of interacting with different types of granular objects has not been exploited. 

TUI has proved to play a vital role in everyday tasks. It makes things more concrete and provides embodiment effects through physicality \cite{baykal2018reviewTUI}.
For example, typical studies such as Project Zanzibar \cite{zanzibar}, which showed a flexible mat to communicate with tangible objects placed on its surface. It also supports sensing a user's touch and hover hand gestures, which opened up the possibility of novel digital experiences. De Tinguy et al. \cite{de2018enhancing} proposed that portable devices could be used to simulate the experience of interacting with different objects. Schmitz et al. \cite{Itsy-Bits} proposed a fabrication pipeline and sensing approach that enabled object recognition of tangibles on capacitive touchscreens. 
Yan et al. use the LaserShoes \cite{yan2023lasershoes} to achieve real-time inference, which cooperate with human under different circumstances. At the same time, they used the laser speckle imaging technique and the LaserShoes could distinguish the surface textures that appear.

However, existing TUI with granular materials only involves a pre-specified type of granular media. 
Ishii et al. \cite{ishii2004bringing} proposed the concept of continuous tangible user interface (Sandscape) using granular materials. They argued that granular materials could bridge the gap between physical and digital forms because of their continuous physical properties. Users could interact with models made of sand and the shape of the granular media was converted into a digital height field to preview land scapes. 
Kazi et al. \cite{sandcanvas} developed a digital canvas (Sandcanvas) to simulate sand drawings. They studied the gestures on touchscreens and incorporated sand simulators to reproduce sand drawings on a virtual screen.
Follmer et al. \cite{jamming} proposed using granular materials to build Jamming user interface. They exploited computer-controlled jamming of granular particles to show its ability in controlling the stiffness of shape-changing objects. In this work, we will introduce dipping as a method to sense different types of granular media and use multiple types of granular media for interactive applications. 
If users would like to fully exploit the interaction experience or map different types of granular objects into the virtual world, existing solutions are not ready for users to interact with different granular objects. One of the missing features is to enable the input device to understand the type of granular media.  

To understand the different types of media, various methods have been developed for object recognition based on vision \cite{Recog,yan2023lasershoes}, inertia data \cite{vibePhone,Viband,viobject}, acoustic signals \cite{SoQr,SweepSense}, or electromagnetic signals \cite{RadarCat,Z-Ring} for various human-computer interactions. Vision-based methods are typical ways for object recognition. However, during the interaction with the granular media, the shape changes of the material, the occlusion or the varying lighting conditions may all affect the recognition results. Furthermore, vision-based solutions require capturing the interaction scene, which may also bring concerns to users in terms of privacy. Haptic recognition is also an important modality for object recognition in the field of computer-human interaction (CHI). Force or tactile signals are used in previous works with machine learning techniques. For example, tactile information has been used for learning the grasp signature for object recognition \cite{sundaram2019learning} or human-environment interactions \cite{luo2021learning}. Wu et al. \cite{Capacitivo} proposed to encode contact information for object recognition on an interactive fabric. 
VibEye \cite{oh2019vibeye}  used the vibration passing through the object to determine the identity of an object. The vibrotactile information received by the finger was represented with a spectrogram and used for object recognition. Researchers also proposed to install proximity sensors on objects to recognize the grasping events \cite{taylor2009graspables}. However, it is costly to customize different objects by embedding sensors inside them.

While humans naturally learn to recognize types of granular material through physical interactions, it is challenging for computers to classify them, especially in user interaction scenarios. Soil, a typical granular medium, is commonly studied in geoscience using the cone penetrometer test (CPT) to ascertain geotechnical properties and identify soil stratigraphy \cite{chen2022study,wisaksono2023soil}. Geologists increasingly employ force data collected from experiments along with machine learning techniques for tasks such as soil spatial mapping, recognition, and classification \cite{ghaderi2019artificial,jong2021state}. However, traditional CPT is costly and time-consuming, requiring heavy and precise tools and strict perpendicular testing directions \cite{lunne2002cone}. In contrast, DipMe provides an inexpensive and easily deployable alternative, allowing for testing with a casual dipping operation, making it more suitable for tangible interactions with particulate media.

Machine learning has revolutionized object recognition in recent years, with techniques such as Time Series classification (TSC) \cite{TSC2017review} utilizing inception modules with a fully convolution network (FCN) \cite{TSC-inception}, a robust temporal feature network (RTFN) \cite{robust-temporal-feature-network}, and a convolution neural network (CNN) \cite{TSC-CNN}. Additionally, researchers have proposed Multivariate Time Series Classification (MTSC) \cite{MTSC2021review}. These machine learning tools have also been leveraged for object recognition in the human-computer interaction community using multimodal signals \cite{zhao2019design,say-and-find-it,SensorNets,Flexel,MechanoBeat}. We propose the use of a more advanced machine learning algorithm, based on the encoder part of the transformer model and multi-channel mechanism, to achieve autonomous recognition of force signal series, surpassing conventional algorithms in performance.

\begin{figure*}[htbp]
  \includegraphics[width=\textwidth]{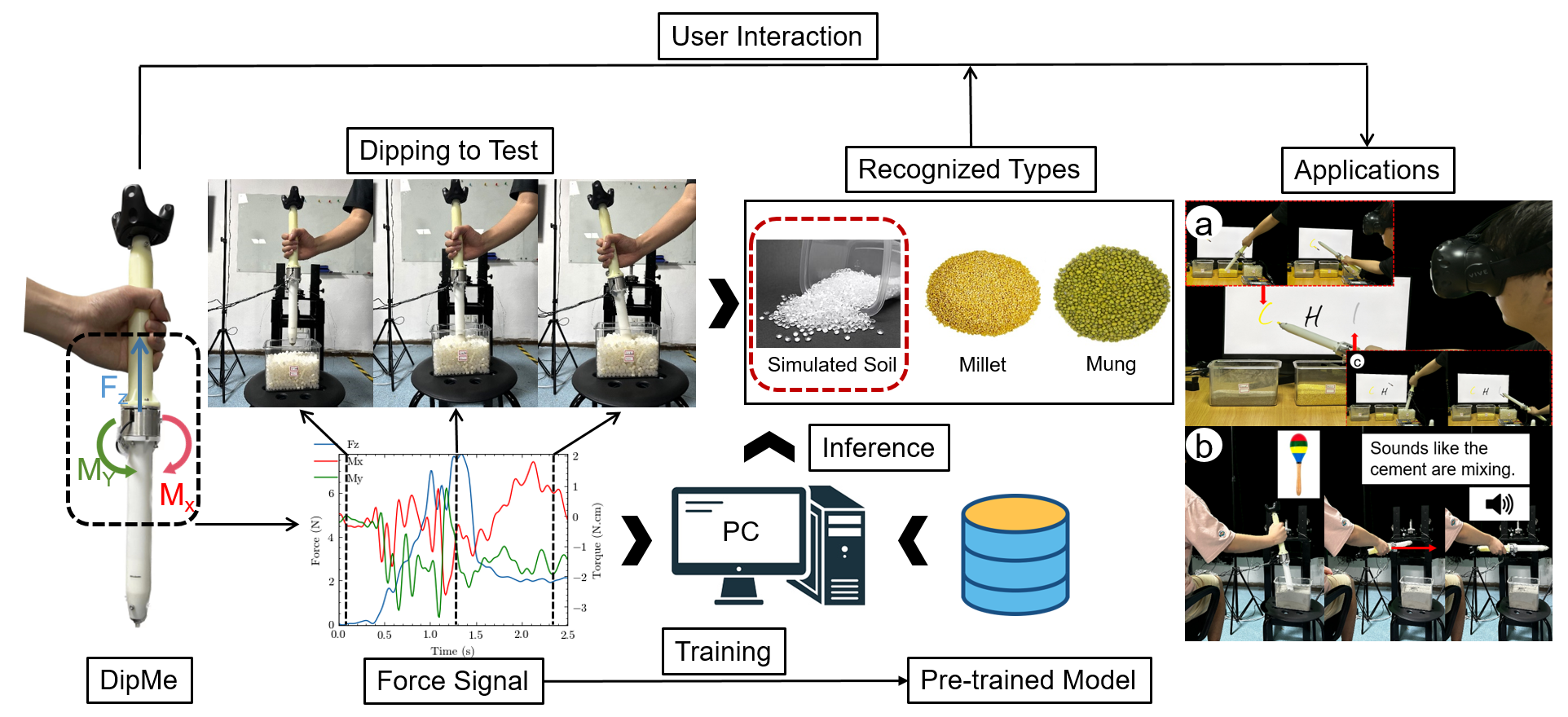}
  \caption{
  DipMe: system overview (left) and applications (right). A user is allowed to dip the device into different granular media. We collect the force and torque signals and use machine learning techniques to recognize the type of the granular material from the multichannel time series data. With the tracked motion of DipMe and the recognized type of granular material, we demonstrate several applications including a virtual drawing interface and a virtual music instrument built with DipMe. 
  }
  \label{fig:teaser}
\end{figure*}
In this work, we propose \emph{DipMe}, a haptic solution for recognizing different types of granular media. Our insight is that humans may probe fingers into the granular particles to feel their haptic experience. Even if they cannot see the granular media, they can still recognize its type based on the force perceived by the fingers. To this end, we develop a device to simulate the probing process and equip the device with force-sensing ability. During the interaction, users are allowed to first probe to test the granular material. We employ modern machine learning techniques to encode the differences of the force signals collected from probing  granular materials. We test our method in various interaction conditions and positive recognition results (92.78\% accuracy with 10 users over 6 granular media) are obtained. With the capability of recognizing types of granular materials, we demonstrate several interactive applications to show the potential of \emph{DipMe} in developing new tangible user interface with granular media.

Our contributions can be summarized as follows:
\begin{itemize}
    \item We propose a new device to intelligently recognize the types of granular material based on haptic information;
    \item We develop new tangible interactive applications that allows user to interact with different types of granular media.
\end{itemize}

\section{System design}\label{sec2}
\subsection{System overview}
Based on the observation from literature, our design of \emph{DipMe} as an input device will consider several aspects. First, the device should be compatible with the probing operation. Second, the device should have the capabilities to acquire the normal force and lateral torque when it probes into the granular media. The force signals should be continuously sampled and the signal-to-noise ratio should be sufficiently large to distinguish the differences when interacting with different granular materials. Lastly, the form of \emph{DipMe} should be close to daily supplies so that it can be easily held and tracked for interactive applications. 

With these design considerations, we prototype a system to recognize granular materials during user interaction, as illustrated in Fig. \ref{fig:teaser}. 
We develop an input device \emph{DipMe} with the force-sensing ability. Users are allowed to hold and dip a pole-shaped device into the granular media. 
\begin{figure*}[htbp]
  \centering
  \includegraphics[width=\linewidth]{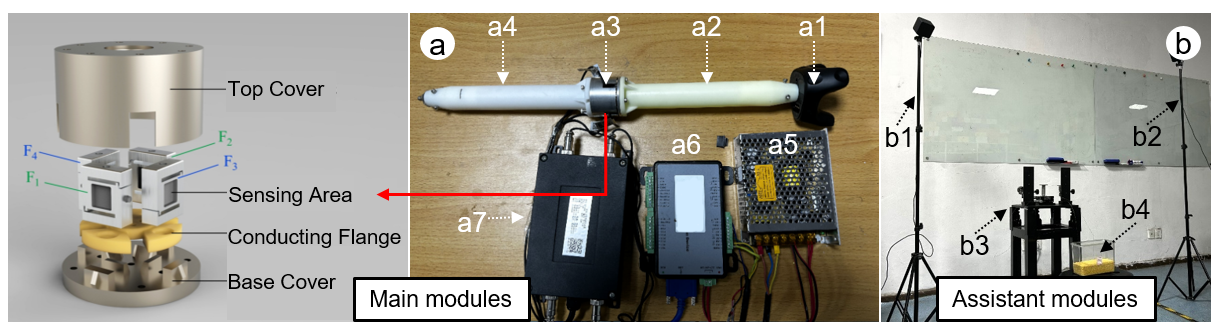}
  \caption{The hardware of measurement system. (a) The main modules of the system, in which a1-a4 make up \emph{DipMe}, a5-a7 actuate the sensors and transmit the collected force signals to PC. a1-HTC ViVE Tracker 3.0, a2-handle side, a3-3D force sensor, a4-sensing side, a5-switching power supply, a6-data acquisition card, a7-signal transmitter. The left subfigure shows the exploded view of the sensor structure (a3): the top cover and conducting flange are made of lightweight and strong PEEK material. The four load cells are arranged in a square formation to form the combined sensing area. (b) The assistant modules of the system, b1 and b2-HTC VIVE Base station 1.0, b3-instrument calibration table, b4-granular media test box.}
  \label{fig:Hardware}
\end{figure*}

With the force and torque signals acquired online, we adopt modern machine learning techniques to solve the multivariate time series classification problem. Along with the recognized types of the granular materials, we use \emph{DipMe} to interact with different granular media. By dipping and operating \emph{DipMe}, we show several tangible interactive applications in interactive mapping of subsurface granular piles, drawing and composing audios by dipping and playing with \emph{DipMe}.
We first present our hardware design and then introduce the computational method for the recognition of granular materials.
\subsection{Hardware design}
The overall form of \emph{DipMe} is designed to share similar structures with soil samplers. We design a pole-like shape to facilitate users to hold and probe \emph{DipMe} into the granular media. 
The entire measurement system consists of five parts: 1) sensing side, which is directly contacted to four load cells of the force sensor, 2) a self-developed three-dimensional force sensor, 3) handle side, which is gripped by users, 4) a VIVE Tracker and 5) assistant modules. The hardware details of \emph{DipMe} are shown in Fig. \ref{fig:Hardware} and different parts are assembled using bolts and nuts.
In our use scenario, a user grasps the upper part (handle side) of \emph{DipMe} and pushes the lower part (sensing side) into granular media with different particle sizes to acquire force signals. A force/torque sensor is installed between the upper and lower part to acquire the force data online. 
In order to obtain the probing motion for both recognition and interaction, we also attach a VIVE Tracker at the top of \emph{DipMe} and send real-time motion data to PC. 

\noindent\textbf{Hardware Configurations. }
The enclosure of the handle side (a2) and sensing side (a4) is 3D printed using nylon and ABS. For contact force (normal force and lateral torque) measurement, we used four small load cells (type: AhJCZN; BCS-M2) from the same batch to form the elastomer part of the combined multidimensional force sensor (a3), each with the same property parameters.
The exploded view is shown in the left subfigure of Fig. \ref{fig:Hardware} . The combined force in the Z-axis results from the sum of the readings from all four cells. The pair of forces detected by the opposing cells constitutes a force couple, which is proportional to the torque in the respective direction. 
Although the torque signals seem to be less prominent than the pressing force, the designed sensor is able to capture the torque with a decent precision. As a difference quantity, the noise of the individual signals will be reduced because the core step of torque measurement is based on the subtraction of the two forces measured at the opposite sides. 
The force sensor is installed between the handle side and the sensing side. The switching power (type: AhJCZN; D-30F, a5) is used to safely supply power to the entire system. The transmitter (type: AhJCZN; LZ-JX4, a7) converts the input from the sensor into an electrical signal and amplifies it for remote measurement and control. All the sensors and other electronic components are connected to a data acquisition card (type: ArtDAQ; USB313XA, a6) with a 100HZ sampling rate. The pose of \emph{DipMe} that can be used to calculate the speed of an object's movement in real time is captured by a tracker (type: HTC; VIVE Tracker 3.0, a1) attached to the top of the handle side with two base stations (type: HTC; VIVE Base station 1.0, b1 and b2). The system error of \emph{DipMe} can be compensated by a calibration table (b3) before each experiment. 

Sample body text. Sample body text. Sample body text. Sample body text. Sample body text. Sample body text. Sample body text. Sample body text.

\section{Methods}\label{sec3}
In this section, we introduce our software solution for haptic recognition of granular media. The captured force and torque signals from the input device serve as input. Initially, we preprocess the raw signals before feeding them into a machine learning pipeline. Subsequently, the recognized label is generated to facilitate interaction with granular media, as depicted in Fig. \ref{fig:teaser}.

\subsection{Data Pre-processing}
\begin{figure*}[!tbp]
  \centering
  \includegraphics[width=\linewidth]{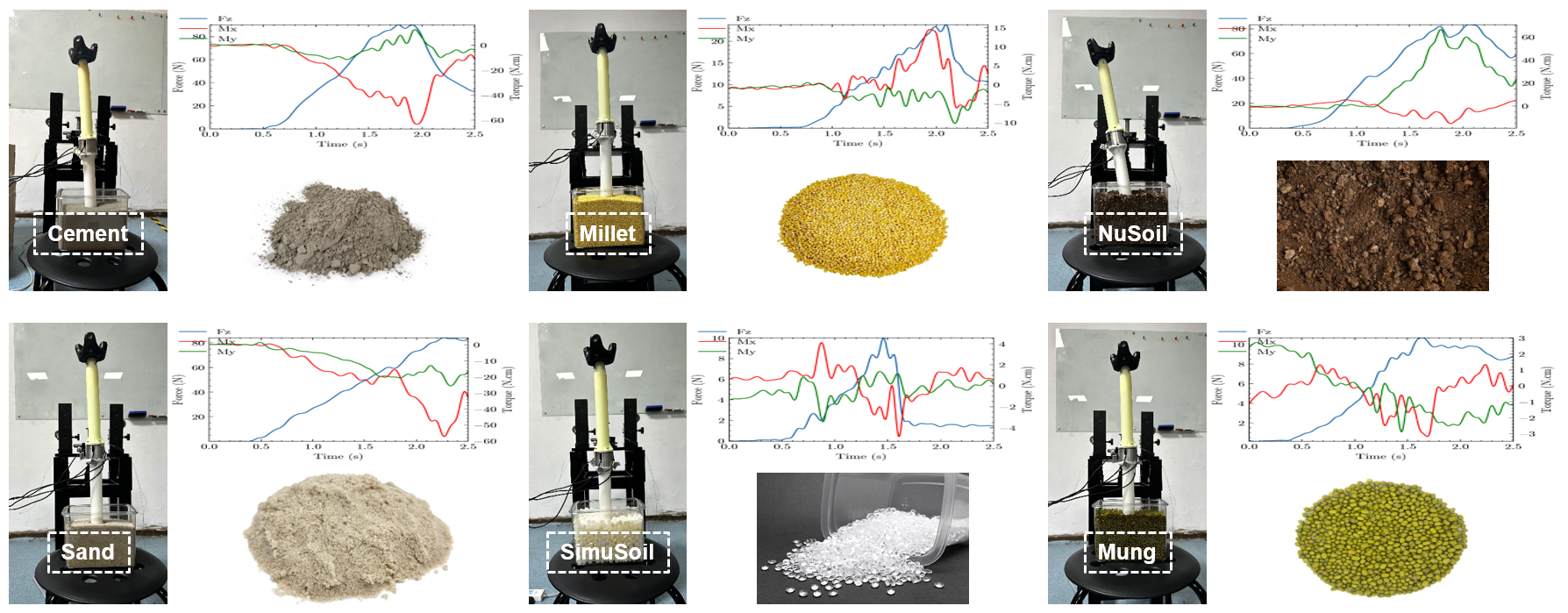}
  \caption{Dipping test experiments and filtered force signals of 6 granular media. Nutrition Soil (NuSoil), whose particle size distribution is about 0$\sim$0.002mm, Millet with 0.007$\sim$0.033mm, Cement with 0.01$\sim$0.02mm \cite{bentz1999effects}, Sand with 0.063$\sim$2mm \cite{kettler2001simplified}, Mung with 3$\sim$4mm and Simulated Soil (SimuSoil) with 7$\sim$8mm.}
  \label{fig:force signal}
\end{figure*}

When a user pushes \emph{DipMe} into the test granular media, the force will be applied to the probe of the sensing side and the signal will be sampled. This I/O force is fed to a series of signal processing methods, which is as shown in Fig. \ref{fig:force signal}.

The processing method for raw signals is as follows. 
First, we convert the gravity of the sensing side to the the base coordinate system (BCS) \cite{thrun2002probabilistic} tracked by the base station. The gravity along the principle direction of \emph{DipMe} as well as the torques contributed by the gravity are subtracted from the raw signals, so that the influence of the orientation of the device will be suppressed. 
\begin{figure*}[htbp]
  \centering
  \includegraphics[width=0.9\textwidth]{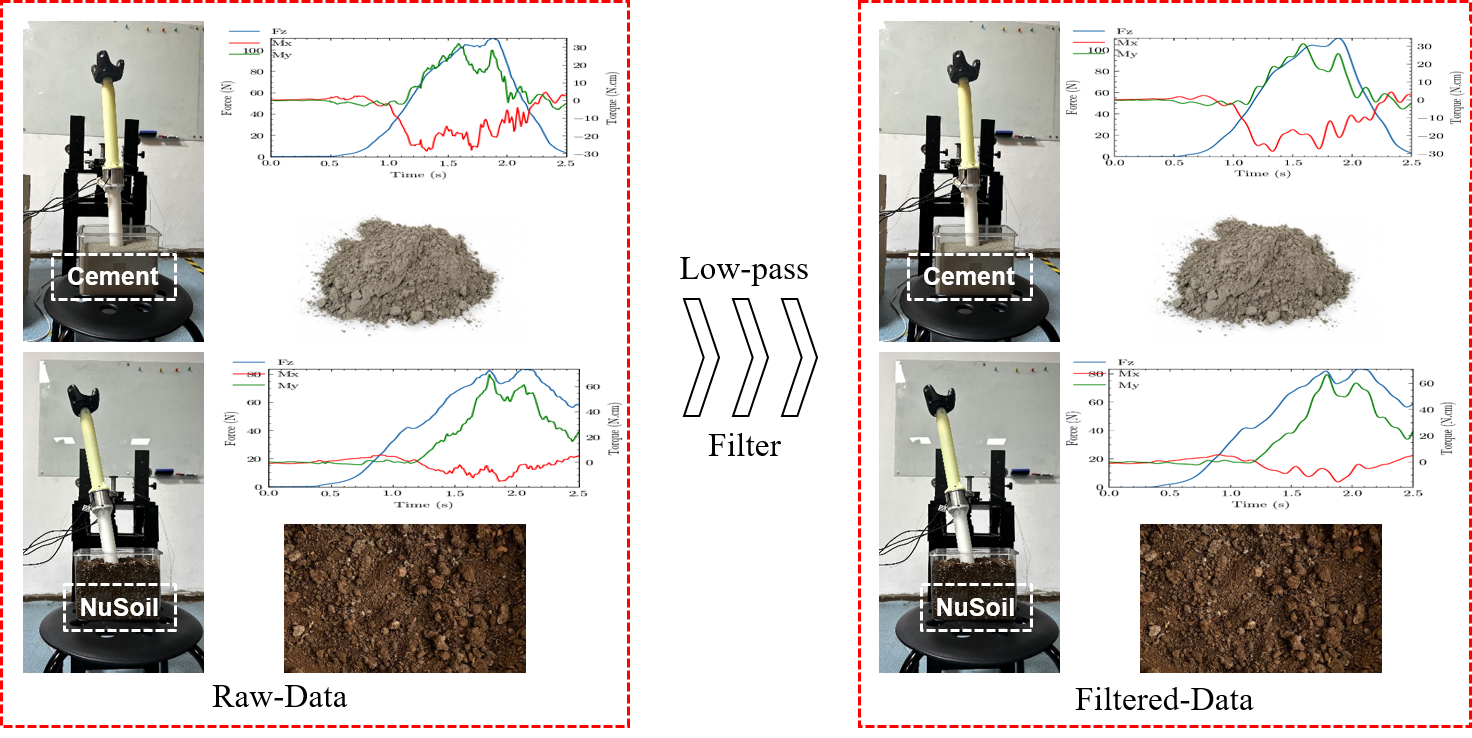}
  \caption{
  Examples of LPF processing.  
  }
  \label{fig:LPF}
\end{figure*}

Second, a low-pass filter (LPF) is applied to eliminate the impact of high-frequency object vibration. The bandwidth of human voluntary motion with wrist motion during daily activities typically falls in the range of around 5 Hz \cite{mann1989frequency}. For LPF, we design an 5-order Butterworth filter that has a flat passband with stopband frequency 10Hz (Fig. \ref{fig:LPF}). 
Finally, we resample the signals according to the velocity extracted from the probing motion, and  unify the data to approximate those collected from a constant-speed probing motion. 
In this case, the force and torque signals are filtered and sensitive to the tested granular media, as shown in Fig. \ref{fig:force signal}.  

\subsection{Multi-channel Encoder Model}
\begin{figure*}[htbp]
  \centering
  \includegraphics[width=0.9\textwidth]{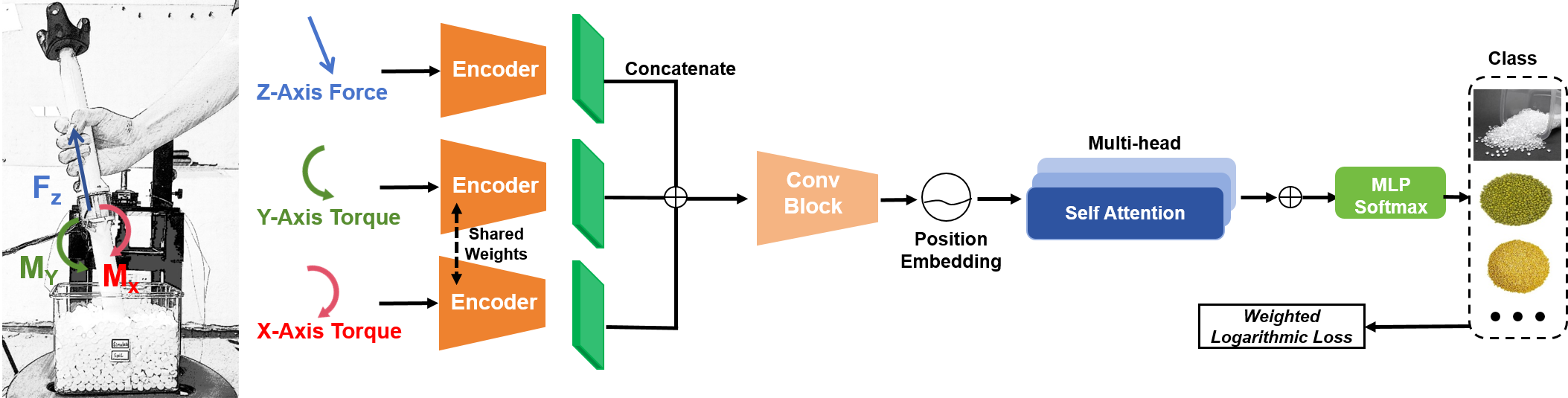}
  \caption{
  Illustration of the Multi-channel Encoder Model.
  }
  \label{fig:Network}
\end{figure*}

Our model is mainly based on the multi-channel encode (MCE) structure with attention mechanism \cite{attention}. In our model, the force signals are first divided into three channels (X-axis torque, Y-axis torque, Z-axis force) and sent to encoding blocks \cite{position-encode}. We then pass them through a ConvBlock and a multi-head attention block (with a position embedding layer, a batch normalization (BN) layer and a full connection (FC) layer) separately. Finally, we concatenate the feature vectors obtained from the three channels into one vector and use a multilayer perceptron (MLP) and a softmax layer to obtain the recognition results, as shown in Fig. \ref{fig:Network}.

A personal computer (PC) with a CPU of 3.10 GHz Inter Core i5-10500 and a GPU of NVIDIA GeForce RTX 3080 is used for doing follow-up experiments. During the training period, we selected Weighted Logarithmic Loss \cite{li2021tactile} as the loss function and applied the Adam optimizer. The learning rate and batch size were set to 3e-4 and 16, respectively. We did not take a pre-trained model to initialize our parameters and used only 100 epochs for model training. In practice, we found that the settings were sufficient for our models to converge. 

\section{Evaluation}
We selected 6 granular media with different particle sizes for our study, as shown in Fig. \ref{fig:force signal}. The single sampling duration is 2.51s
, which is sufficiently long for accurate recognition of the types of granular media. These media could be classified into four groups based on particle size: 1) clay (0$\sim$0.002mm), 2) silt (0.002$\sim$0.063mm), 3) sand (0.063$\sim$2mm), 4) gravel (2$\sim$63mm) \cite{kroetsch2008particle}. For each type, we used a transparent crisper (MBL; MBL-3500, b3) with a volume of 1800mL to store the samples, experimented 40 times with \emph{DipMe}, and then processed the raw sensor data as described in Section 5.1. This step resulted in a time-series matrix (40 times*6 types*251 length*3 channels), and they were used as input signals to granular media recognition.

\subsection{Recognition Length Selection}
As presented in Section 5.2, the model's input is the multivariate-time series with dimension of 3 and length of $N$ from the data pre-processing phase, and the output is the type of granular media. The Recognition Length is set to 128 in our granular media classification. To determine the effectiveness of this length, we use the sliding window approach \cite{wei2006semi} to segment the processed time series data with different lengths to train a set of MCE models (200 samples for training, 40 samples for testing). The tested Recognition Lengths included 32, 64, 128, and 251. 
\begin{table*}
\centering
  \caption{The mean classification accuracy and inference time for single recognition with various recognition length.}
  \label{tab:recog len select}
  \begin{tabular}{c|cccc}
    \toprule
    Recognition Length &32&64&128&251\\
    \midrule
    $Accuracy$ &77.33\%&83.14\%&94.37\%&\textbf{95.51\%}\\
    Inference Time &\textbf{0.41$s$}&0.80$s$&1.44$s$&2.88$s$\\
  \bottomrule
\end{tabular}
\end{table*}
Table \ref{tab:recog len select} presents the mean accuracy and the inference time (includes a sampling time) for the classification of one trial. As anticipated, input series with larger length result in greater accuracy, yet they require significantly longer time for classification. Given that the accuracy improvement is moderate, we have opted to choose 128 as the recognition length for input series in our model. This decision has been taken while striking a balance between accuracy and inference time for the online recognition task.
\begin{table*}
\centering
  \caption{Evaluation metrics of different methods using 10-fold validation.}
  \label{tab:model compare}
  \begin{tabular}{c|ccccc}
    \toprule
    Method &gRFS&DTW+KNN&MCNN&ResNet&MCE\\
    \midrule
    $Accuracy$ &70.15\%&81.88\%&85.65\%&89.70\%&\textbf{94.85\%}\\
    $Precision_{macro}$ &0.752&0.831&0.873&0.925&\textbf{0.981}\\
    $Recall_{macro}$ &0.702&0.824&0.881&0.899&\textbf{0.957}\\
    $F1$-$Score_{macro}$ &0.726&0.827&0.877&0.912&\textbf{0.969}\\
  \bottomrule
\end{tabular}
\end{table*}
\subsection{Method Comparison}
We ran cross-validation test \cite{wong2015performance,jung2018multiple} using four popular MTSC 
approaches (Generalized random shapelet forest (gRFS) \cite{karlsson2016generalized}; Dynamic Time Warping with K-Nearest Neighbor (DTW+KNN) \cite{tran2019novel}; Multi-scale Convolutional Neural Network (MCNN) \cite{cui2016multi}; ResNet \cite{wang2017time}) to demonstrate the effectiveness of the proposed method (MCE with the recognition length 128) for 10-fold cross validation.

In order to better compare the different methods, we selected four classification metrics ($Accuracy$, $Precision_{macro}$, $Recall_{macro}$, $F1$-$Score_{macro}$\footnote{Macro-value represents the average of values across all categories, and this indicator remains unaffected by data imbalances.}) \cite{berger2020threshold,lewis1996training} as evaluation metrics and the results are provided in Table \ref{tab:model compare}. 
With both the self-attention mechanism and the multi-channel fusion mechanism, our method effectively identified patterns within the multivariate time series with a higher performance than that of other MTSC methods. 
\section{User Study}
We conducted a user study to assess the effectiveness of \emph{DipMe}, for recognizing the types of granular media with actual users. We recruited 10 right-handed participants (8 males and 2 females), with ages ranging from 21 to 28 years old (mean = 23.3, SD = 2.31) for this user study via volunteering. During the study, the participants held the handle side of \emph{DipMe} with their dominant hands and they were told to move the \emph{DipMe} casually and try to maintain a normal speed when testing the granular material. 
\subsection{Experiment Details}
Our first experiment was to test whether the model could recognize the types of granular media when different participants conducted the tests with \emph{DipMe}. To ensure there is no overlapping between the training set and the testing set, we first divided all the data (including the data in the Section 5) into seven folders and randomly selected three folders (180 samples) as the testing data set while ensuring a consistent number of samples in each type. The second experiment was to do cross-user validation using the leave-one-out evaluation method \cite{wong2015performance}. We retrained the model using data from 8 participants (144 samples) and tested it on the remaining participants (36 samples).

All participants finished the two sessions, first with a short teaching and practice session, which can familiarize participants with the entire system and provide them with information on the experimental procedure (the time duration is 1$\sim$2min and practical data were not included in data set). For each type of the granular media, the second session was repeated for 3 times (the time duration is 3$\sim$4min). We randomly order the placement of granular media and calibrated to eliminate the influence of \emph{DipMe}'s gravity before each repetition. During the experiment, participants followed the instructions displayed on the host computer. In total, we collected 180 samples on 6 kinds of granular media from the 10 participants, and it took less than 1 hour to finish all the data collection. All data will be normalized prior to training.

\subsection{Results}
\begin{table*}
\centering
  \caption{Accuracy of the trained model and cross-user model on six types of granular media}
  \label{tab:two experiments accuracy}
  \begin{tabular}{c|cccccc}
    \toprule
    Experiment &Nutrition soil&cement&Sand&Millet&Mung&Simulated soil\\
    \midrule
    Trained-model &86.67\%&\textbf{90\%}&\textbf{90\%}&\textbf{96.67\%}&\textbf{100\%}&\textbf{96.67\%}\\
    Cross-user-model &\textbf{100\%}&83.33\%&66.67\%&83.33\%&83.33\%&83.33\%\\
  \bottomrule
\end{tabular}
\end{table*}

\begin{figure*}[!tbp]
  \centering
  \includegraphics[width=\linewidth]{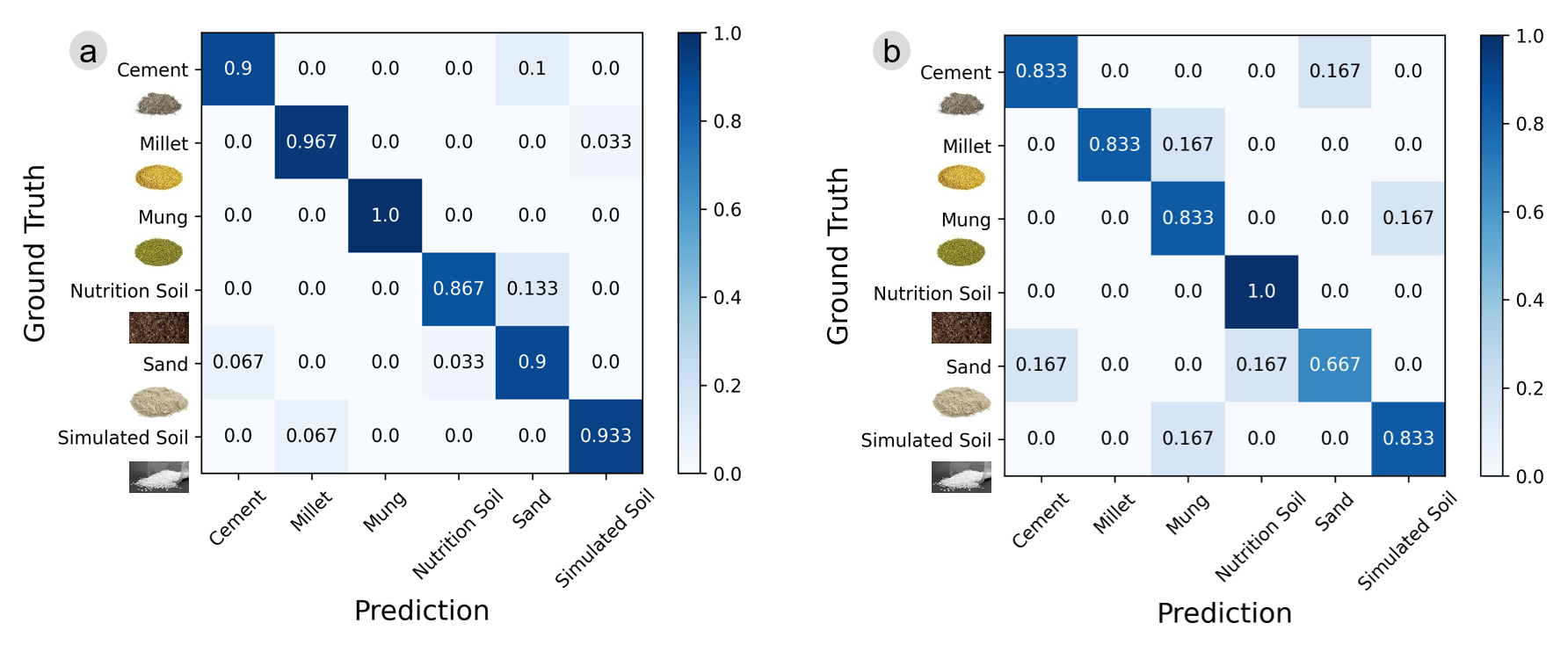}
  \caption{The confusion matrices of the 6 granular media from two experiment models. (a) Recognition results using a trained model with 240 training samples and 180 testing samples. (b) Recognition results using a cross-user model with 144 training samples and 36 testing samples.}
  \label{fig:confusion matrix}
\end{figure*}

A confusion matrix \cite{townsend1971theoretical} for two experiments is shown in Fig. \ref{fig:confusion matrix}, where a decimal in the (i,j) cell signifies the percentage of the i-th media recognised as the j-th media. It is visible that the diagonal terms are dominant for almost all types.

\noindent\textbf{Recognition Accuracy of the Types of Granular Media with Actual Users. } The average accuracy over all granular media and participants was 92.78\% (Fig. \ref{fig:confusion matrix} (a)), with the macro\footnote{The average recall equals the average accuracy since the number of presenting each media was all the same.}-precision of 0.932 (SD = 0.072), the macro-recall of 0.928 (SD = 0.049), the macro-F1 score of 0.929 (SD = 0.052). This result substantiates the efficacy of our approach in recognizing the types of granular media penetrated by different participants. 

\begin{figure*}[!tbp]
  \centering
  \includegraphics[width=\linewidth]{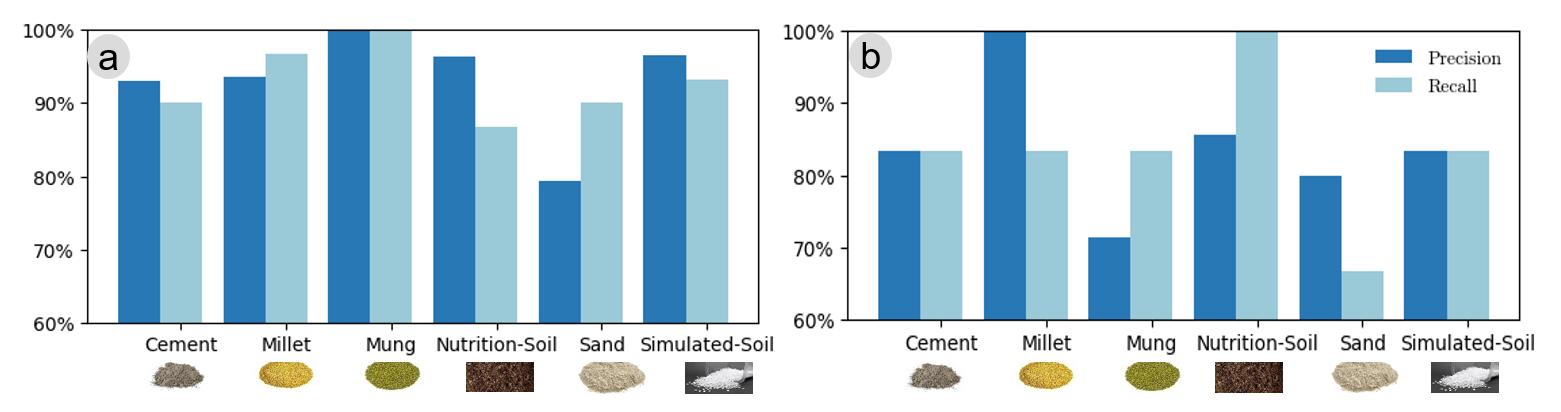}
  \caption{The precision and recall bar graphs of two experiment models of the 6 granular media. (a) Precision and Recall using a trained model with 240 training samples and 180 testing samples. (b) Precision and Recall using a cross-user model with 144 training samples and 36 testing samples.}
  \label{fig:precision and recall}
\end{figure*}

The most significant errors occurred with the nutrition soil, which was misidentified as sand in 13.3\% of cases (Fig. \ref{fig:precision and recall} (a)). Nutrition soil is particularly prone to moisture-related changes during extended experimental cycles, lasting approximately 2-3 weeks. This can lead to some of the soil samples used in the experiment acquiring mechanical properties similar to those of sand and will cause some recognition errors in our model. Individual precision and recall values are shown in Figure \ref{fig:precision and recall} (a) for all the granular media. These statistics reaffirm the effectiveness of \emph{DipMe}.

\noindent\textbf{Cross-User Validation Results. } For cross-user validation, the recognition results are shown in Fig \ref{fig:confusion matrix} (b). The average recognition accuracy of the retrained cross-user model is 83.32\%. From the Fig. \ref{fig:precision and recall} (b) we can calculate that, the macro-precision is 0.840 (SD = 0.093), the macro-recall is 0.833 (SD = 0.105), the macro-F1 score is 0.832 (SD = 0.076). 
The average accuracy decreases during cross-user validation. The decrease in accuracy may be attributed to variations in force patterns resulting from differences in penetration motion pattern \cite{escobar2013dynamic}, leading to discrepancies between the training and testing datasets. It's important to note that the same type of granular media can correspond to multiple  force patterns, contributing to reduced accuracy in cross-user validation.

The results also indicate that simulated soil and mung exhibit the higher accuracy among the six media with varying characteristics (see Table. \ref{tab:two experiments accuracy}). This is understandable as the structure of simulated soil and mung is simpler, leading to more distinguishable force patterns. Additionally, sand has the lowest average accuracy. We can also find that there is a decrease of the performance between trained-model and cross-user-model. One possible reason for this observation is that media with varying particle sizes had an amplifying effect on the differences in participants' probing motion, resulting in greater disparities in force patterns within the same type of media groups.

\begin{figure*}[htbp]
  \centering
  \includegraphics[width=0.8\linewidth]{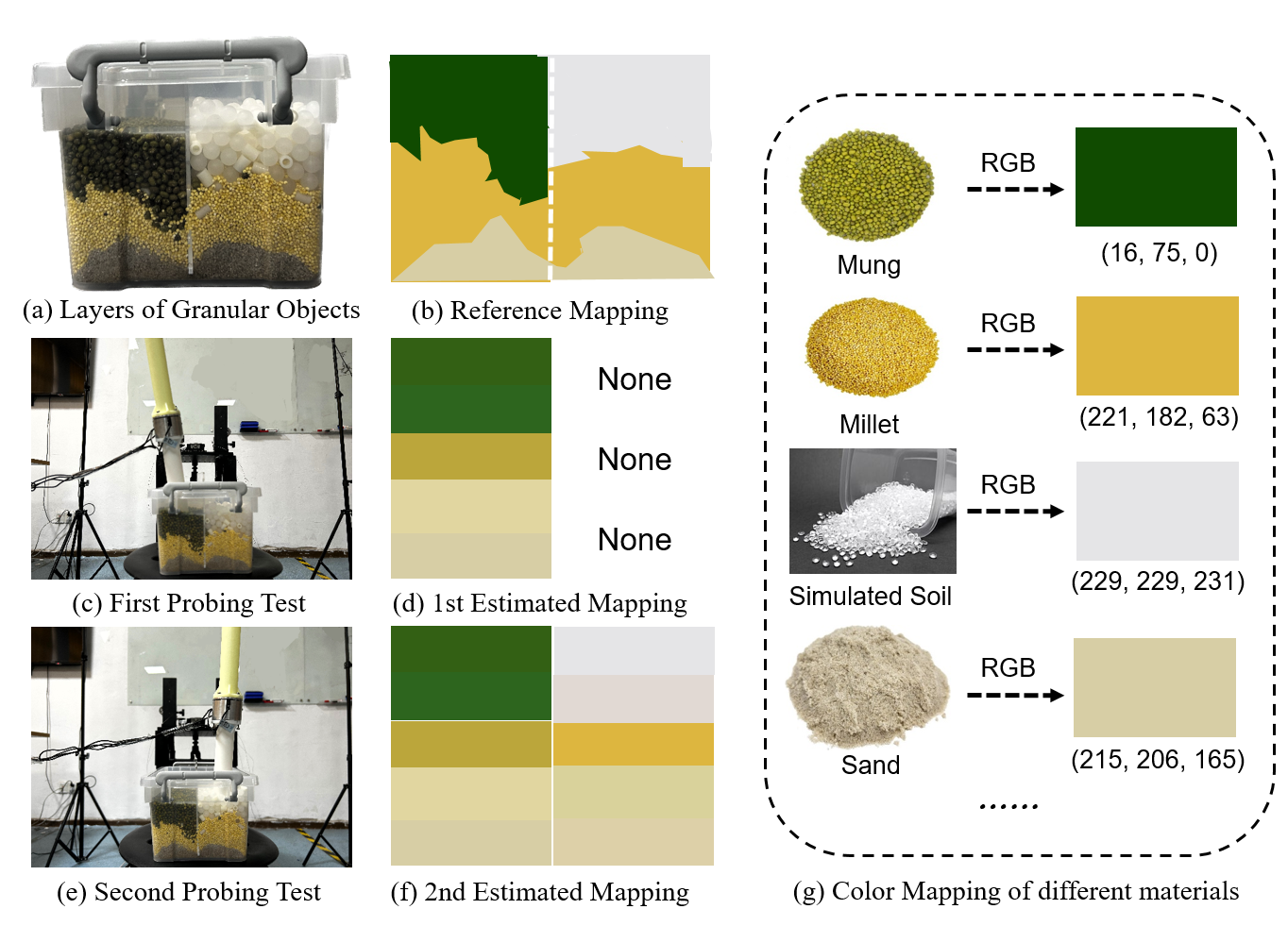}
  \caption{A quick and rough spatial mapping of granular media using \emph{DipMe}. 
  (a) four types of granular materials are piled in a lab box (the left side is arranged in layers ordered as mung-millet-sand, while the right side simulated soil-millet-mung). (b) The reference distribution of the granular materials. (c,e) two probing test at the left and right sides. (d,f) The estimated subsurface material mapping, which is visualized by mixing the color maps in (g) by the predicted probabilities estimated by \emph{DipMe}.
  }
  \label{fig:application 1}
\end{figure*}

\section{Application}
\begin{figure*}[htbp]
  \centering
  \includegraphics[width=0.9\linewidth]{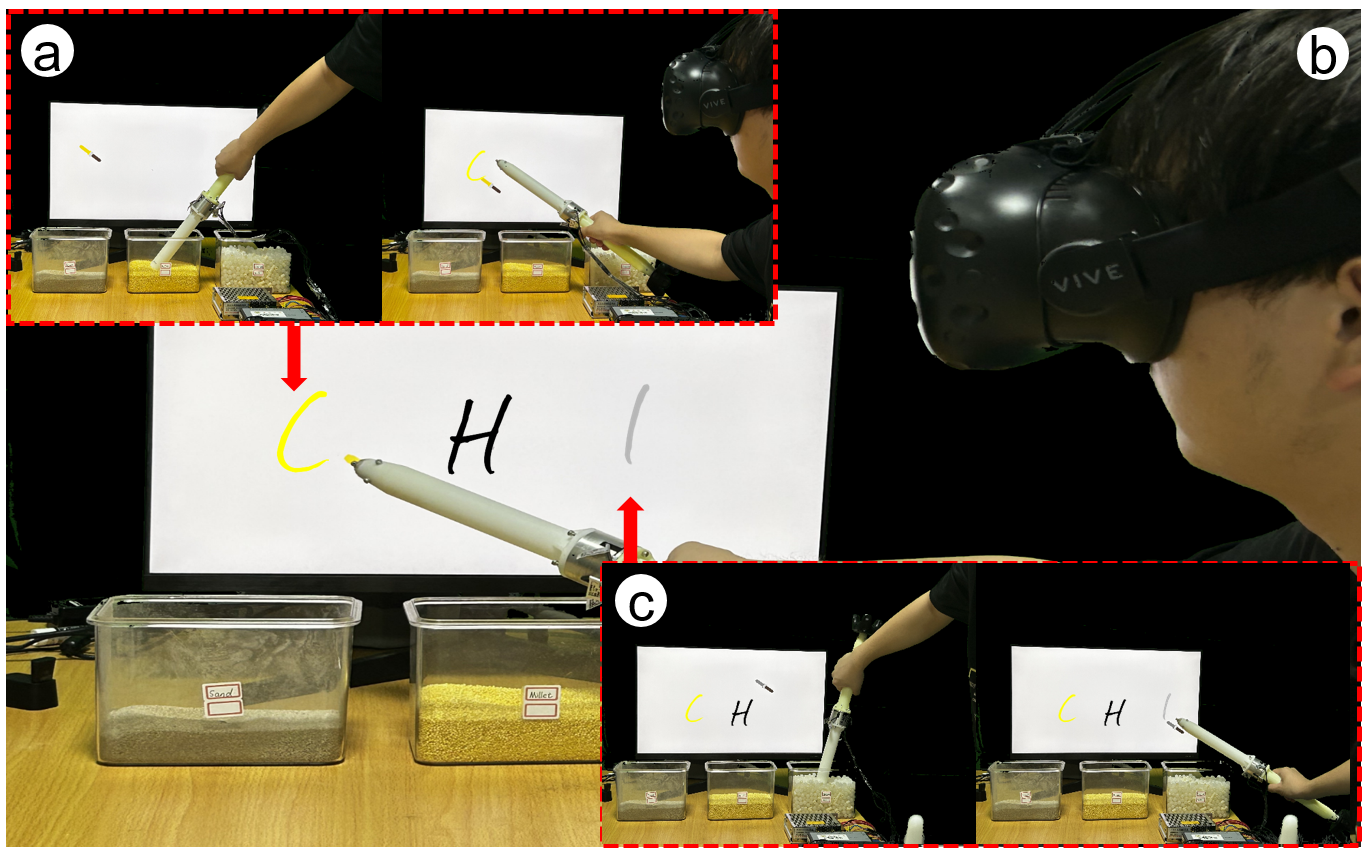}
  \caption{The application of \emph{DipMe} in a drawing interface (b). After dipping into the millet, the system recognized its type, assigning the virtual brush with a yellow color, and the user proceeded to write the letter C. (c) The system recognized Simulated Soil, turning the virtual brush grey, and the user proceeded to write the letter I.}
  \label{fig:application 2}
\end{figure*}

\emph{DipMe} enables users to identify various types of granular materials through tangible props, resulting in a straightforward and immersive tangible interaction experience. Following HCI community's design principles, we have designed three applications as shown in Fig. \ref{fig:application 1}-\ref{fig:application 3}.

\subsection{Spatial Mapping Interface of Subsurface Granular Media}
In geotechnical engineering \cite{ghaderi2019artificial}, a common task is to map the spatial distribution of various subsurface soil types. With the material recognition capability of \emph{DipMe}, we propose to dip and estimate a map of the subsurface granular material. 
Unlike the traditional systems for spatial mapping of soil, which are time-consuming and costly, \emph{DipMe} provides users with a natural interface to get a quick and rough map of subsurface granular material types. 

For this application, we dipped the proposed input device into the soil with layers of different granular materials, as shown in Fig. \ref{fig:application 1} (a,c,e). With the tracked pose of \emph{DipMe} and the predicted probability of the granular media, we built a map of the subsurface soil as visualized in Fig. \ref{fig:application 1} (d,f). 
We used different colors to represent different types of granular media and applied alpha compositing to represent the predicted probability of each type of granular material. We probed \emph{DipMe} into the soil from different locations, sampled each probing operation with five nodes, mixed the predicted probability and mapped them to colors. The results shown in Fig. \ref{fig:application 1} (d,f) illustrated a rough map of the subsurface structures, which well aligned with the soil. Note that the subsurface information was not visible to users and we took a transparent container to help illustrate the results. If the number of experiments on the same longitudinal profile increased, the resolution of the resulting spatial maps would become higher. 
Although the accurate mapping of the subsurface material distribution requires expensive devices such as computer tomography equipment, we regard the results from \emph{DipMe} as a map of the haptic cues to help understand the subsurface structure quickly with a low cost. 

\subsection{Drawing Interface for VR}
\emph{DipMe} can be used to build a tangible interactive application for drawing and writing. As shown in Fig. \ref{fig:application 2}, a user employed \emph{DipMe} to recognize the type of granular media provided, with a gesture similar to a brush dipped in different colors of ink in the real world. He could then complete the switching of virtual brush colors and proceed to subsequent writing tasks in space. We equipped \emph{DipMe} with a virtual reality (VR) interface as the input device. By tracking the 6D tracker (HTC; VIVE Tracker 3.0) and showing the writing pattern on the screen or see-through HMD, users were allowed to draw and update the property of the strokes by dipping into the granular media. With the passive haptic provided by the dipping operation, users are enabled to feel the update of the stroke properties by perceiving the friction and texture of different granular media. 

\subsection{Shaking Virtual Maracas with Changeable Tones}
\begin{figure*}[htbp]
  \centering
  \includegraphics[width=\linewidth]{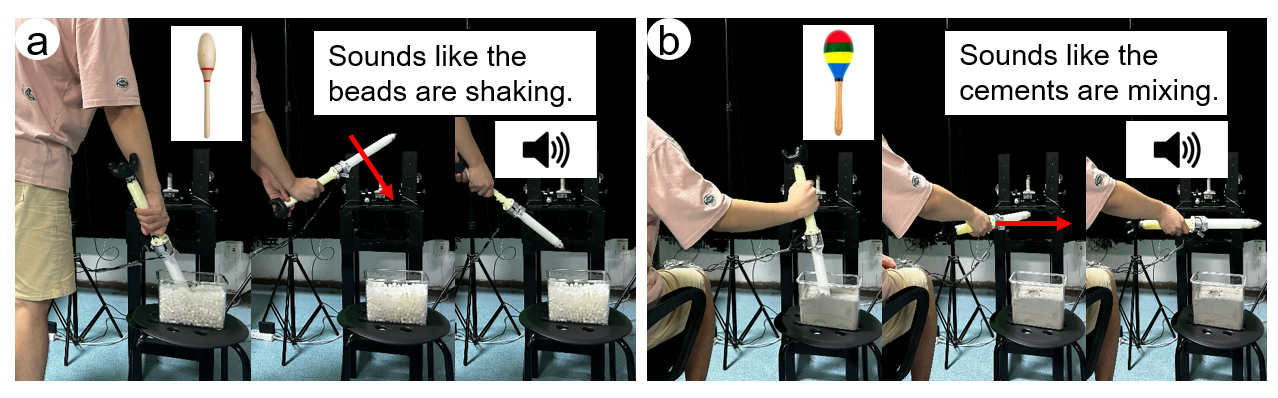}
  \caption{
  The application of \emph{DipMe} as a virtual maraca. The user dipped the device into beads (a) and cement (b), \emph{DipMe} recognized the type of the granular material and the system played similar tones as if the user was shaking a maraca filled with beads or cement. 
 }
  \label{fig:application 3}
\end{figure*}

As shown in Fig. \ref{fig:application 3}, we developed an interface to simulate Maracas\footnote{A maraca is a percussion instrument commonly found in Caribbean and Latin American music genres.} using \emph{DipMe}. In the physical world, shaking a maraca filled with different materials will produce varying frequencies of sound, while changing the tones of the instrument needs to switch to another maraca with different designs. We propose to simulate Maraca with \emph{DipMe} and an interface to change its tone by dipping into various granular media. As depicted in Fig. \ref{fig:application 3}, a user probed \emph{DipMe} into one granular medium. The system recognized the type of the granular media and equipped \emph{DipMe} with a similar sound from shaking a maraca filled with the granular particles. By naturally switching the instrument with a dipping gesture, the user could compose a clip of music by shaking one \emph{DipMe} to simulate virtual maracas with changeable tones. 

\section{Discussion}
\noindent\textbf{Probing Motion. } In this work, the recognition of the granular material is based on the haptic data collected from a probing operation. In our applications, a casual dipping motion works well in the interaction. However, if the granular material is too firm, e.g., cement flour will be hard to dip especially when it gets wet, we may need to twist or shake the device while probing it. As the model is trained from the signals generated by straightly probing the device into the granular media, if the user would like to shake or twist the device to probe it, we need to classify those special motions and retrain the model. Nevertheless, if the straight probing motion is allowed, our model still works when the motion is operated by different individuals, as validated in Section 7. Therefore, the recognition from active dipping is an easy operation for recognizing granular material.  

\noindent\textbf{Latency. } As for the probing operation, we require the signals to be sufficiently long for object recognition. In our implementation, we need to take the force and torque signals whose length is about 1.28 seconds. Because the online inference stage of the learning model is fast (less than 0.2 seconds in our system), the latency for obtaining the recognition results is also about 1.48 seconds, similar to other tangible user interfaces with object recognition as the preprocess \cite{oh2019vibeye}. In our test, we find the latency does not break the user experience, while we will also work to see if we can reduce the latency of the model in the future.  

\noindent\textbf{Multimodal. } One possible solution to improve the sensing unit of the input device is to add the information from more channels. For example, the vibration sensed from accelerometers or the sound recorded when the device is dipped into the granular media. We also note that for the simplicity of the design and the privacy of the interaction, we still recommend to use the haptic information. Data from other modes are not quite suitable for the interaction, e.g., the vibration will be easily affected by the signature of the user's motion, the acoustic data might be polluted by the sound in the open environment or the vision data will be expensive to capture and may reveal too much private information during the interaction. In our practice, the haptic recognition as well as the dip motion is a good balance of the design and work well in our tangible applications.    

\noindent\textbf{More Tangible Applications. } We show several new tangible interactive applications with granular material in this work enabled by \emph{DipMe}. In order to fully exploit the benefits of tangible user interfaces, we may equip the recognized type of granular media with more simulators in the virtual environment to provide rich visual-haptic feedback to the users. For example, we may equip \emph{DipMe} with SandCanvas \cite{sandcanvas}, so that we can generate sand drawing on different granular particles with the system intelligently perceive the type of the granular media. In this case, we can also share the visual feedback from the simulated particle animation along with the passive haptics from interacting with the real-world granular media. 

\noindent\textbf{Miniature Design. } In this work, we implement \emph{DipMe} with its size close to a handheld pole. The demonstrated interactions also fit with holding a large brush or maraca. If we change the size of the force sensor to be smaller, we can further reduce the size of \emph{DipMe} for pen-style or even finger-wearable forms and enable new dipping experience. Therefore, a miniature design of the hardware is also one of the directions for the future work.

\section{Conclusion}
We propose \emph{DipMe}, a new input device for haptic recognition of different types of granular media. We show that the haptic signal from a dipping operation is a valid feature to help computer understand types of granular media with modern machine learning tools. The recognition performance during user interactions is evaluated and several applications enabled by \emph{DipMe} are demonstrated to show its potential in developing new tangible interfaces with the proposed input device. 

Our work is a first step towards the recognition of granular media types for human-computer interaction. In the future, we will explore how to design and utilize other information such as tactile signals that are sensitive to more properties of the granular material, such as humidity or finegrained friction. Furthermore, we will work on the miniature design of the input device and explore more potential applications by integrating \emph{DipMe} into VR or AR systems. 







\begin{appendices}

\section{Development details of the Force/Torque Sensor}\label{secA1}

Here we briefly describe the development details of the force/torque sensor introduced in the paper. As shown in the left subfigure of Fig. \ref{fig:Hardware}, we employed four compact one-dimensional force sensors as the load cells. Each load cell has its measuring range of $\pm 40 N$ and its accuracy is $\leq 0.3\% F.S.$. By composing the readings from the cells, we obtain the force along the z-axis and the torques along the x-axis and y-axis. Given the symmetric layout of the loading cells, the composed force and torques are computed as
\begin{equation}\nonumber
\label{eq1}
\left\{
\begin{aligned}
F_z &= F_1+F_2+F_3+F_4  \\
M_x &= k_x(F_3-F_4)\\  
M_y &= k_y(F_2-F_1)   
\end{aligned}
\right.
\end{equation}
where $F_i$ is the force reading of the $i$th cell, $k_x$, $k_y$ denote the torque coefficients which are related to the distance between the opposite loading cells. Considering the fabrication errors, friction or other factors which may affect the measurements, we determine $k_x$, $k_y$ and add other compensation terms by calibrating the sensor with standard loads.  We obtain the performance parameters of the developed sensor and list them in the following table. 
\begin{table*}[h]
\centering
\caption{Performance Parameters of the Sensor Developed for \emph{DipMe}}
\label{tab:sensor parameters}
\begin{tabular}{c|c}
\toprule
Indices &  Parameters\\
\midrule
Measuring Range of $F_z$ & $0 N \sim 100 N$ \\
Accuracy of $F_z$ & $1.5\% F.S.$ \\
\midrule
Measuring Range of $M_x$ & $-1 Nm \sim 1 Nm$ \\
Accuracy of $M_x$ & $2\% F.S.$ \\
\midrule
Measuring Range of $M_y$ & $-1 Nm \sim 1 Nm$ \\
Accuracy of $M_y$ & $2\% F.S.$ \\
\midrule
Size &Diameter of $50 mm$, height of $30 mm$\\
\bottomrule
\end{tabular}
\end{table*}



\end{appendices}
\bigskip

\bibliography{sn-bibliography}

\end{document}